\documentclass{DISproc}

\begin{document}
% ------------------------------------
\title{New features in version 2 of the fastNLO project}

% for single authors the superscripts are optional
\author{{\slshape Daniel Britzger$^1$\footnotetext{Talk given by D. Britzger at the XXth International Workshop on Deep-Inelastic Scattering (DIS12), University of Bonn, March 2012}, Klaus Rabbertz$^2$, Fred Stober$^2$ and Markus Wobisch$^3$}\\
  (The fastNLO Collaboration)\\[1ex]
  $^1$DESY, Hamburg, Germany\\
  $^2$KIT, Karlsruhe, Germany\\
  $^3$Louisiana Tech University, Ruston, Louisiana, USA}

% please enter the contribution ID for the DOI
\contribID{xy}

\doi % if there is an online version we will register DOIs

\maketitle

\begin{abstract}
  Standard methods for higher-order calculations of QCD cross sections
  in hadron-induced collisions are time-consuming. The fastNLO project
  uses multi-dimensional interpolation techniques to convert the
  convolutions of perturbative coefficients with parton distribution
  functions and the strong coupling into simple products.  By
  integrating the perturbative coefficients for a given observable
  with interpolation kernels, fastNLO can store the results of the
  time-consuming folding integrals in tables, which subsequently are
  used for very fast rederivations of the same observable for
  arbitrary parton distribution functions, different scale choices, or
  $\alpha_s(M_Z)$.  Various tables with code for their evaluation are
  available for numerous jet measurements at the LHC, the TeVatron,
  and HERA\@. FastNLO is used in publications of experimental
  results by the ATLAS, CMS, CDF, D0, and H1 collaborations, and in
  all recent PDF analyses by ABM, CTEQ, HERAPDF, MSTW and NNPDF\@. This
  article focuses on developments implemented in the new version~2 of
  fastNLO, enhancing and broadening its functionality.
\end{abstract}

\section{Introduction and the fastNLO concept}

Precision measurements in high energy physics reveal their full power
only if they are compared to accurate theoretical predictions.
% Precision measurements in high energy physics can be exploited to their full power
% only if they are compared to accurate theoretical predictions.
For the interpretation of experimental data or the extraction or tuning of
model parameters reasonably fast theory calculations are needed.  For
measurements defined in published cross sections, repeated
computations of (almost) the same cross sections have to be performed.
Typical examples where calculations have to be repeated more often for
the same observable are:
\begin{itemize}
\item Comparisons of data vs.\ theory, where e.g.\
  different sets of parton distribution functions (PDFs) are used,
  like those provided by the global fitting collaborations.
\item Determination of the PDF uncertainty on theory predictions.  As
  an example, the uncertainty on the cross sections arising from the
  NNPDF~\cite{Ball:2011mu} PDF set require $100$ to $1000$
  rederivations of the same cross sections where only the PDFs
  differ.
  % PDF4LHC recommendation \cite{Alekhin:2011sk}
\item Fitting of PDFs where for each step in the iterative procedure
  the cross sections have to be recalculated for the corresponding
  temporary PDF\@.  FastNLO is used by various global PDF fitting
  groups, like MSTW~\cite{Martin:2009iq}, CTEQ~\cite{Lai:2010vv},
  NNPDF~\cite{Ball:2011mu}, ABM~\cite{Alekhin:2012ig}, or
  HERAPDF~\cite{Nowak:12}.
\item Determination of model parameters, or the determination of the
  strong coupling constant in iterative fits.
\item Determination of the theory uncertainty of cross sections from
  missing higher orders, where conventionally the calculations are
  repeated for different choices of the renormalization and/or
  factorization scale.
\item Studies of the scale dependence of the theory cross sections.
  For processes involving multiple scales it is not clear, which scale
  setting to choose.
  %%% For example in $pp$ dijet measurements, reasonable choices for
  %%% the renormalization/factorization scale would be the average
  %%% transverse momentum of the two leading jets, the maximum
  %%% transverse momentum of the leading jet or some formula including
  %%% the rapidity distance between the two leading jets and their
  %%% invariant mass.
\end{itemize}

Some observables in next-to-leading order (NLO) or even higher-order
calculations can be computed rather fast like e.g.\ DIS structure
functions.  Other observables like Drell-Yan and jet cross sections,
however, are very slow to compute. Especially the latter ones are the
current focus of fastNLO\@. FastNLO provides computer code and
precalculated tables of perturbative coefficients for various
observables in hadron-induced processes.
% This allows the very fast computation of these observables for
% arbitrary PDFs and $\alpha_s(M_Z)$ and also the opportunity to
% change the definition of the scales.
The calculation of the fundamental cross sections or matrix elements is
performed by flexible computer code like
\texttt{NLOJet++}~\cite{Nagy:1998bb,Nagy:2001xb,Nagy:2001fj,Nagy:2003tz},
which has been used here.

For illustrating the ideas in this document, all formulae are shown for
jet production cross sections in deep-inelastic scattering.
All concepts and formulae can be generalized to
hadron-hadron collisions like at the LHC or the TeVatron~\cite{Kluge:2006xs}.

% nlojet\\
% alle pdf gruppen verwenden fastNLO (mit zitat)\\
% lots of calcualtion\\

\subsection{The fastNLO concept}

Perturbative QCD predictions for observables in hadron-induced
processes depend on the strong coupling constant $\alpha_s$ and on the
PDFs of the hadron(s). Any cross section in lepton-hadron or
hadron-hadron collisions can be written as the multiplication of the
strong coupling constant to the power of $n$, $\alpha_s^n$ , the
perturbative coefficients $c_{i,n}$ for the partonic subprocess $i$,
and the corresponding linear combination of PDFs from the one or two
hadrons $f_i$, which is a function of the fractional hadron momenta
$x_a$, $x_b$ carried by the respective partons, as:
\begin{equation}\label{eq:CrossSection}
  \sigma_{ep\rightarrow\mathrm{jets}}(\mu_r,\mu_f) =
  \sum\limits_{i,n}\int\limits_0^1dx\alpha_s^n(\mu_r)c_{i,n}(\frac{x_{Bj}}{x},\mu_r,\mu_f)f_i(x,\mu_f)
\end{equation}

The calculation of this cross section with reasonable statistical
precision is very slow due to the necessary Monte Carlo integration of
the accessible phase space.  The idea of fastNLO is to separate
the PDFs and the $\alpha_s$ factors from the perturbative coefficients
$c_{i,n}$ and to convert this integration into a
sum~\cite{Pascaud:1994vx,Wobisch:00}.  This discretization introduces
a set of eigenfunctions $E_i(x)$ (with $\sum_i E_i(x) \equiv 1$) around
a defined number of $x$-values.  The PDF in eq.~\ref{eq:CrossSection}
can then be replaced by $f_i \simeq \sum_i f_a(x_i) E_i(x)$ and is
removed from the integral.  When calculating the perturbative
coefficients the nodes receive fractional contributions of each event
within the $x$-range of each eigenfunction. The perturbative
coefficients are calculated once with very high statistical precision
and are stored in a \emph{table}.  The remaining integration over $x$
to compute the cross section is turned into a sum over the $n$
perturbative orders, $i$ parton flavors, and all the $x$-nodes.  To
respect the scale dependence, a similar procedure is employed.
The multiplication of the PDF, $\alpha_s$,
and the perturbative factors are performed when reading the table of
all $c_{i,n}$ values, which is very fast and gives the opportunity to
change the PDFs and $\alpha_s$ as required.  When calculating
hadron-hadron cross sections, the two $x$ integrations %over the two partons
are replaced by two sums instead of only one like in the
depicted DIS case.

% hadron hadron collisions -> Reference Markus.

\section{New features in fastNLO version~2}

Here, new features of the fastNLO version~2 are presented in
comparison to the previous version~1.4~\cite{Kluge:2006xs}.
The fastNLO project comprises three main elements, the \emph{table}
format, \emph{creator} code to create and \emph{reader} code to read
and evaluate the tables with an interface to PDFs.\\
A new \emph{table} format features more flexibility and foresees to
incorporate multiple additive or multiplicative contributions to the
cross sections.  Threshold corrections, which were already available
previously for hadron-hadron induced inclusive jet
production~\cite{Kidonakis:2000gi}, higher-order calculations,
electroweak corrections, or new physics contributions can be
implemented in a similar way as soon as they are available.  Further
multiplicative correction factors like non-perturbative corrections
can be stored together with their uncertainties.
Also data can be included within the new table format together with
their correlated and uncorrelated uncertainties.\\
The computation of the tables has been optimized.  An automated scan
to determine the covered $x$-range is performed first.  The now
flexible number of $x$-nodes for each analysis bin respects their
different $x$-coverage.
The scale dependence is stored as a separate array and also the
interpolation of the scale nodes is optimized.\\
The fastNLO \emph{reader} code for evaluating the fastNLO tables is
now available in Fortran as well as in independently developed
\texttt{C++} classes with an agreement of $\mathcal{O}(10^{-10})$
between the two.  It is distributed in one
package~\cite{Fnlo:2012:Online} as open source code, which is
installable following the GNU autotools procedure with the only
dependence on some external functionality to access PDFs like in
LHAPDF~\cite{Whalley:2005nh}.  Previously released tables keep their
validity and can be converted into the new format.

% Both packages are distributed without any further dependencies
% (like \texttt{CERNLIB} or \texttt{ROOT}), although some PDF
% interface
% must be provided (typically \texttt{LHAPDF} or \texttt{QCDNUM}).\\

% arbitrary number of dimensions for binning\\
% easy to install: autotools\\

% neues format ist ultra präzise! -> plot

% \begin{figure}[htb]
%   \centering
%   \includegraphics[width=0.45\textwidth]{britzger_daniel_fastnlo_fig1}
%   \includegraphics[width=0.45\textwidth]{britzger_daniel_fastnlo_fig2}
%   \caption{ Caption Todo }
%   \label{Fig:Precision}
% \end{figure}

\section{The generalized concept of flexible-scale tables}

A generalized concept for even more flexible tables was also released
in fastNLO version~2.
These are called \emph{flexible-scale} tables and are based on two principles.\\
The dependence on the renormalization and factorization scales
$\mu_{r/f}$ can be factorized when calculating the
perturbative coefficients $c_{i,n}$, like
\begin{equation}\label{eq:ScaleIndependentWeights}
  c_{i,n}(\mu_r,\mu_f) = c_{i,n}^0 + \log(\mu_r)c_{i,n}^r +  \log(\mu_f) c_{i,n}^f.
\end{equation}
This way, only scale independent weights $c^0$, $c^r$, and $c^f$ are
stored in three scale independent fastNLO tables.  The multiplication
of the scale dependent $\log$ terms are performed only when evaluating
the table.  Similarly to the $\alpha_s$ term in
eq.~\ref{eq:CrossSection}, where $a_s(\mu_r)$ can be regarded as an
arbitrary function of the scale $\mu_r$, also the scales $\mu_r$ and
$\mu_f$ can be regarded as functions of any relevant $k$ observables
$s_k$, like $\mu_{r/f} = \mu_{r/f}(s_1,s_2)$.
FastNLO examples employ $k=2$ since each observable needs a separate
interpolation array and increases the required evaluation time.\\
This method gives the opportunity to store multiple possible scale
definitions, like e.g.\ the jet momentum $p_\mathrm{T}$ and the event
virtuality $Q^2$.  When evaluating the fastNLO table it is now
possible to choose any function of these two scale settings for the
definition of the renormalization and factorization scale. Typical
examples are e.g.\ $\mu_r^2 = (Q^2+p^2_\mathrm{T})/2$ and $\mu_f^2 =
Q^2$, but also definitions like $\mu = s_1 \cdot \exp(0.3 \cdot s_2)$
are possible.  Further, the scales can be varied independently and by
arbitrary scale factors.  This gives new opportunities to study the
scale dependence of cross sections.  The concept is also available for
$pp$ and $p\overline{p}$ calculations.  For future applications, the
\emph{flexible-scale} concept is valid also for higher-order
calculations like NNLO without any significant loss in speed.

% scale independent weights\\
% scale independent weights\\
% three tables for the weights\\
% three (any number of) look-up tables\\
% log terms are calculated during evaluation time\\
% mu is regarded as function\\
%
% 1) We can choose μR independently from μF\\
% 2) We can choose the functional form of μR/F as functions of look-up-variables\\
% \\
% Choose scale composition from previously stored scales\\
% Also scale variation for μr and μf are thus independently possible through\\
% New options for scans of scale dependence\\

% ATLAS case\\
% ATLAS Dijet Invariant Mass, r=0.6\\
% pT and y* are stored in table\\
% Ren./fact. scale can be any function of (pT,y*)\\
% fastNLO vs. plain NLOJet++ calculation with free choice of ren./fac. scale\\

% Precision ~ $\mathcal{O}10^-3$\\

\section{Showcase application}

\begin{figure}[htb]
  \centering
  \includegraphics[width=0.98\textwidth]{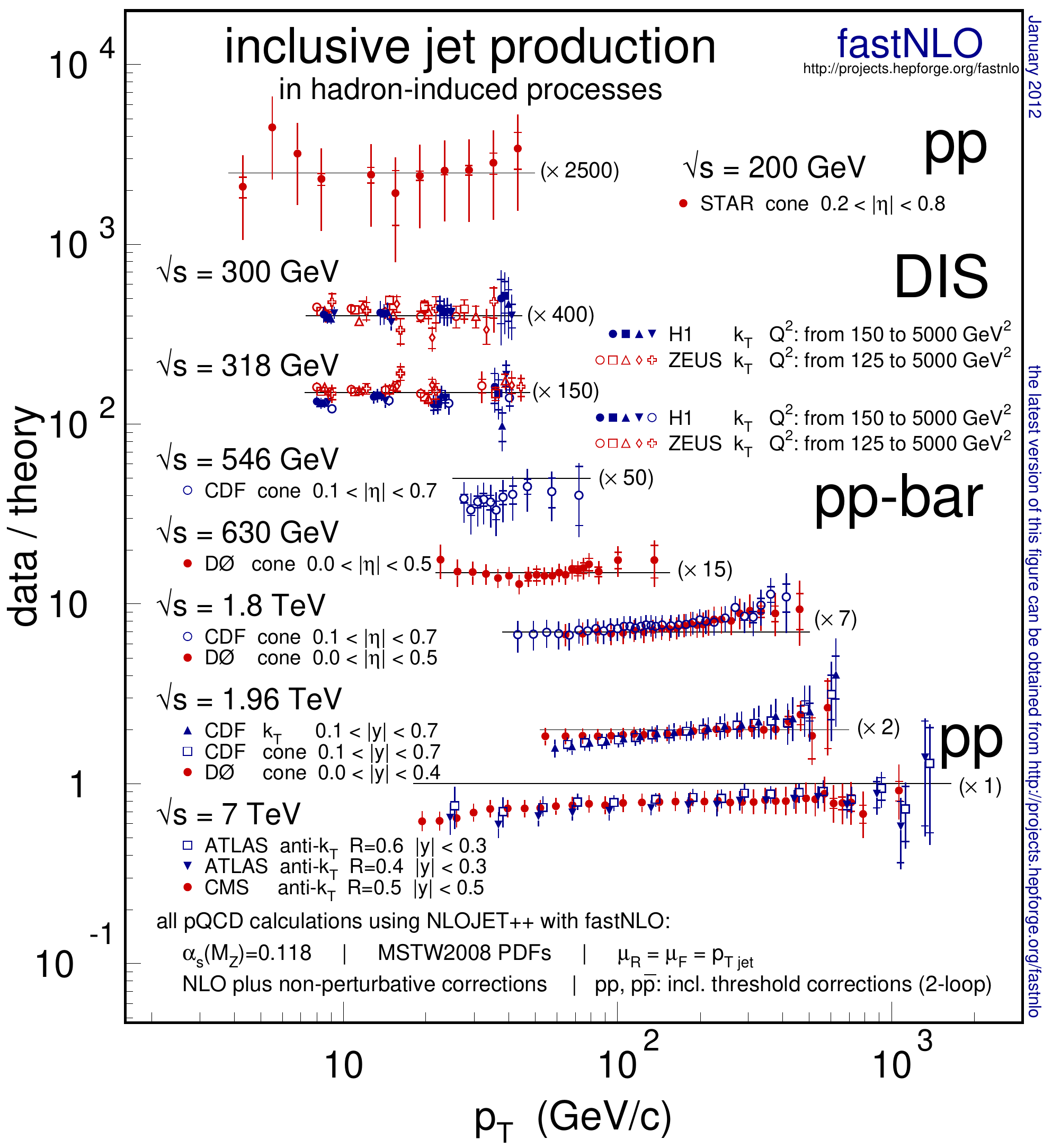}
  \caption{Ratios of data and theory for inclusive jet cross section
    measured in hadron-hadron collisions and in deeply inelastic
    scattering at different center-of-mass energies. The ratios are
    shown as a function of jet transverse momentum $p_\mathrm{T}$. The
    theory results are computed for MSTW2008 PDFs.}
  \label{Fig:Alljets}
\end{figure}

Numerous fastNLO tables are available on the fastNLO website
\cite{Fnlo:2012:Online} for various measurements by ATLAS, CDF, CMS,
D$\emptyset$, H1, STAR, and ZEUS\@.  All calculations were performed
using the \texttt{NLOJet++} program~\cite{Nagy:1998bb,Nagy:2001xb,Nagy:2001fj,Nagy:2003tz},
for calculating the matrix elements.  A data/theory comparison of global
inclusive jet data in hadron induced processes as a function of the
transverse jet momentum for various center-of-mass energies is shown
in fig.~\ref{Fig:Alljets} employing the MSTW2008 PDF
sets. Hadron-hadron induced processes further include $2$-loop
threshold corrections, which represent a part of NNLO\@. An
updated version of this plot is available through the arXiv
article~\cite{Wobisch:2011ij}.

% ****************************************************************************
% BIBLIOGRAPHY AREA
% ****************************************************************************

{\raggedright
  \begin{footnotesize}
    % IF YOU DO NOT USE BIBTEX, USE THE FOLLOWING SAMPLE SCHEME FOR
    % THE REFERENCES
    % ----------------------------------------------------------------------------
    % ----------------------------------------------------------------------------

    % IF YOU USE BIBTEX, - DELETE THE TEXT BETWEEN THE TWO ABOVE
    % DASHED LINES - UNCOMMENT THE NEXT TWO LINES AND REPLACE
    % 'smith_joe.bib' WITH YOUR FILE(S)

    \bibliographystyle{DISproc} \bibliography{britzger_daniel_fastnlo}
  \end{footnotesize}
}

% ****************************************************************************
% END OF BIBLIOGRAPHY AREA
% ****************************************************************************

\end{document}